# Metastable atomic layer deposition: 3D self-assembly towards ultra-dark materials

**Authors:** Mario Ziegler[1,2*], André Dathe[3], Kilian Pollok[4], Falko Langenhorst[4,5], Uwe Hübner[1], Dong Wang[2] and Peter Schaaf[2]

**Affiliations:**

[1]Competence Center for Micro- and Nanotechnologies, Leibniz Institute of Photonic Technology Jena (IPHT), Albert-Einstein-Straße 9, 07745 Jena, Germany

[2]Chair of Materials for Electrical Engineering and Electronics, Institute of Materials Science and Engineering and Institute of Micro- and Nanotechnologies MacroNano®, TU Ilmenau, Gustav-Kirchhoff-Str. 5, 98693 Ilmenau, Germany.

[3]Single-molecule Microscopy Group, Jena University Hospital, Friedrich Schiller University, 07745 Jena, Germany

[4]Institute of Geosciences, Friedrich Schiller University Jena, Carl-Zeiss-Promenade 10, 07745 Jena, Germany

[5]Hawaiʻi Institute of Geophysics and Planetology, School of Ocean and Earth Sciences and Technology, University of Hawaiʻi at Manoa, Honolulu, HI 96822, USA

*Correspondence to: mario.ziegler@leibniz-ipht.de

## Abstract

Black body materials prove promising candidates to meet future energy demands as they are able to harvest energy from the total bandwidth of solar radiation. Here, we report on high absorption (> 98 %) near-black body-like structures consisting of a silica scaffold and Ag nanoparticles with a layer thickness below 10 µm; fabricated using metastable atomic layer deposition (MS-ALD) and to be applied for a wide solar spectrum ranging from 220 nm to 2500 nm. Several effects contribute collectively and in a synergistic manner to the high absorbance, including the pronounced heterogeneity of the nanoparticles in size and shape, particle plasmon hybridization and the trapping of omni-directionally scattered light in the 3D hierarchical hybrid structures. We propose that, in the future, MS-ALD needs to be considered as a simple and promising method to fabricate black-body materials with excellent broadband absorption.

**Manuscript**

Clever combinations of different materials can lead to material setups with superior optical properties that cannot be achieved with their individual components, only.[1] Going one step further, the combination of geometries of different size scales can create a completely new category of material setups coming with superior properties, such as a negative refractive index or broadband absorption.[2,3] These so-called metamaterials contain sub-wavelength structures which facilitate the creation of artificial permittivity and permeability. This range of possibilities in the design of optical properties leads to an unique optical performance and fosters applications such as nanoruler[4,5], cloaking[6–8] or negative refractive index engineering.[2,9–11]

Taking into account the absorption bandwidth, metamaterials can be fabricated reliably and with high performance in the microwave regime,[6,7] while they suffer from rather narrow frequency ranges in the visible regime.[5,6,12,13] In 2015, Fratalocchi et al. [14] overcame this issue by introducing their so-called "dark chameleon dimers" system which absorbed 98-99 % of the incident light (400 nm to 1400 nm) at a layer thickness of 10.2 µm, independent of polarization or incident angle. In comparison, commercially available carbon nanotubes absorb only 70-85 % at similar thicknesses. Fratalocchi et al. synthesized the structures by the seeded growth of gold nanospheres from gold nanorods. Here, each structure consists of individual nanorods (length: 75 nm ± 7 nm, diameter: 18 nm ± 2 nm) with one randomly attached nanosphere (30 nm ± 3 nm).

In this paper, we report on a broadband absorber structure for the visible (VIS) and near-infrared (NIR) region, following a recently developed ALD fabrication protocol known as metastable atomic layer deposition (MS-ALD).[15,16] The resulting structures are highly absorbent from the VIS to the NIR region (> 98 % in the range of 220 nm to 2500 nm). The high broadband absorbance stems from the complex optical behaviors in the self-assembled 3D hierarchical silica structures loaded with silver (Ag) nanoparticles (Fig. 1(a)). Usually, 3D nanostructures fabricated with ALD require pre-patterned 3D templates that are conformally coated by ALD.[17,18] In contrast to classical ALD, for MS-ALD, the deposition parameters of the ALD process are altered to exploit the metastability of a noble metal template. With this, 3D broadband absorber structures can be generated without the need for complex 3D pre-patterned structures. For the fabrication of our broadband absorbers, we referred to the most promising structures from our previous work.[16,19] Thus, for this study, we used Ag films of 300 nm thickness as the template in the silica deposition process with precursors of tri(dimethylamino)silane (TDMAS) and oxygen plasma. The silver oxidizes to silver oxide, when oxygen plasma is present and becomes metastabile at elevated temperatures above 100°C.[15] Metastable silver oxide will dissolve again under the release of reactive oxygen. Side reactions for the formation of excess silica occur with the released oxygen from the cyclic decomposition of the metastable silver oxide, leading to the evolution of the desired 3D structures in a self-assembled way. This two-step process (Ag deposition, MS-ALD synthesis of 3D structures) we refer to as a "supercycle". More details are given in the supplementaries.

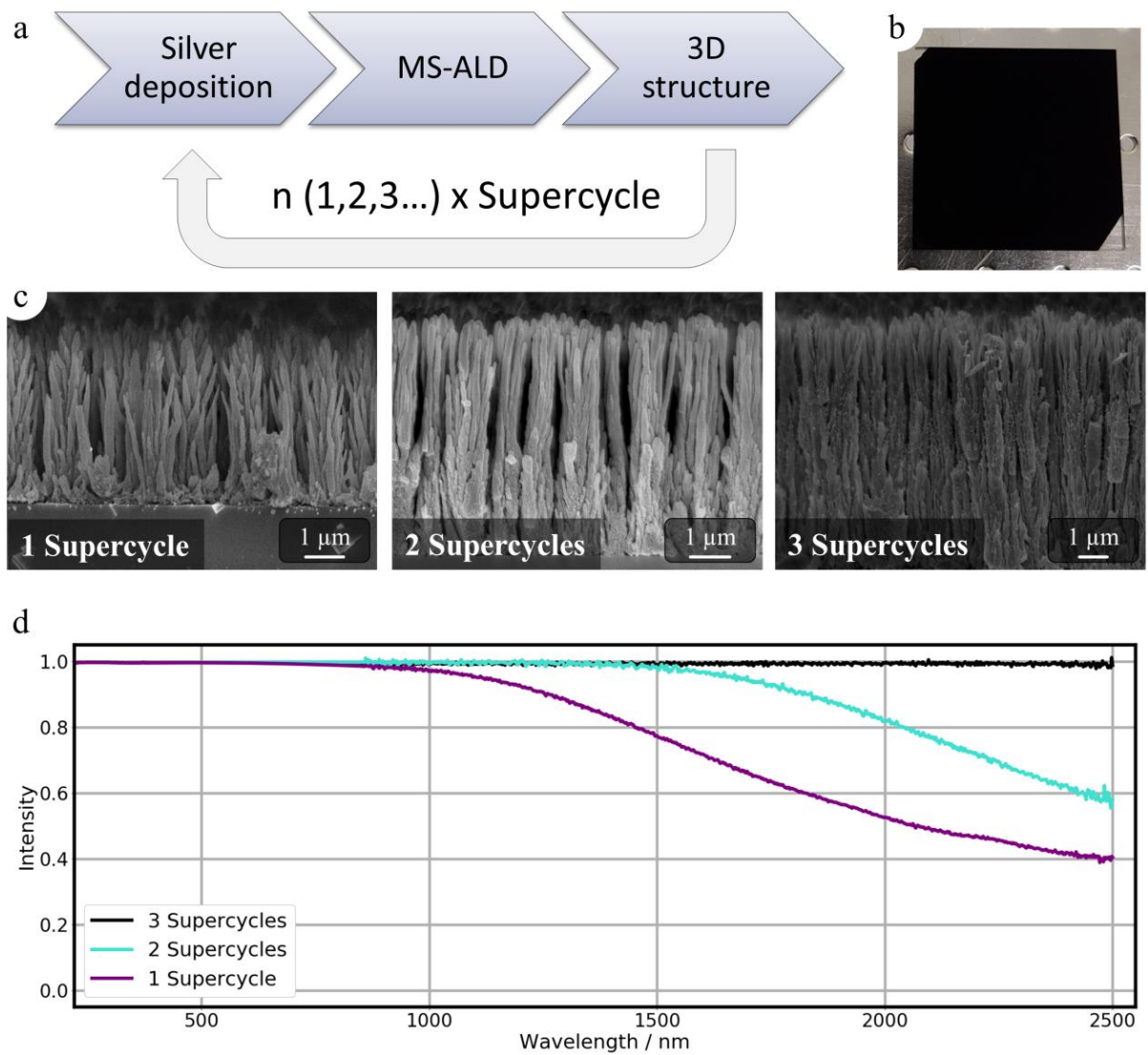

**Fig. 1: Overview on the absorption behavior of the Ag-silica hybrid structures. (a) Concept of MS-ALD supercycles to generate multilayered absorber structures. (b) Optical image of a 2.5 × 2.5 cm² glass sample coated with the Ag-silica hybrid nanostructure. (c) Cross-section SEM images of absorber structures fabricated with 1, 2 and 3 supercycles. (d) Extinction spectra of a sample as shown in (b) and (c), for one MS-ALD supercycle (1×), two MS-ALD supercycles (2×) and three MS-ALD supercycles (3×) in the wavelength range from 220 nm to 2500 nm, TDMAS dose time: 50 ms, oxygen plasma dose time: 6 s, Ag film thickness: 300 nm.**

Fig. 1(b) shows a photograph of the broadband absorber fabricated on glass substrate with MS-ALD. The corresponding SEM image is shown in Fig. 1(c). The nanostructures are completely matte black and distributed homogeneously over the whole substrate surface. They are 4.2 µm in height and consist of silica structures in an herb-like assembly, covered with randomly distributed Ag nanoparticles. The diameters of the single nanowires are in the range of 145 nm ± 116 nm. Fig. 1(d) depicts the absorption, A, for the wavelength range from 220 nm to 2500 nm which was calculated by the relation $A = 1 - T - R$. As can be seen from the absorption graph of the MS-ALD-structure (purple), the sample shows a broadband absorption of 99 % from the initial 220 nm to 760 nm. At longer wavelengths, the absorption gradually decreases and has its minimum of 40.4 % at 2500 nm. The transmission gradually increases from approx. 0 at 220 nm to 1 % at 820 nm. In the NIR region, the transmission increases significantly up to its highest value of 44.4 % at 2500 nm. The reflectance slightly rises from 1228 nm in the NIR region and is 15.4 % at 2500 nm. In the same plot, the absorbance of the structure fabricated using 2 MS-ALD supercycles is presented in (turquoise). Here, the previously generated 3D structure was again coated with an Ag film of 300 nm which then transformed to a complex 3D architecture (Fig. 1(a)). As can be seen, the absorbance increases remarkably, exhibiting a very high absorption (> 99 %) from 220 nm up to 1400 nm, which is comparable to black-body materials made of disordered nanostructures.[14] Both transmission and reflection (see supplementary information) decrease to 10 % intensity in comparison to the single supercycle. The absorption range could be enhanced even more by simply applying a third supercycle in addition to the dual supercycle system. Structures generated from 3 supercycles revealed an absorbance of almost 98 % over the complete measurement range. The structure height also increased further to 9.0 µm ± 0.2 µm. Measurements did not reveal any quantifiable diffuse transmission or diffuse reflection. A comparison of all three structures and the corresponding optical measurements are given in the supplementary information (see Fig. S1).

Several studies predict that the high absorption behavior in the visible region would be caused by the interaction of the incorporated Ag nanoparticles with the incident light, so-called "localized surface plasmons".[20–23] However, localized surface plasmons resonance (LSPR) of an individual and small (<30 nm) Ag nanoparticle results in a rather narrow absorption peak, if calculated using the Mie solutions.[24] The peak position and its absorption intensity are functions of the nanoparticle material, surrounding media, particle shape and size.[20,25,26]

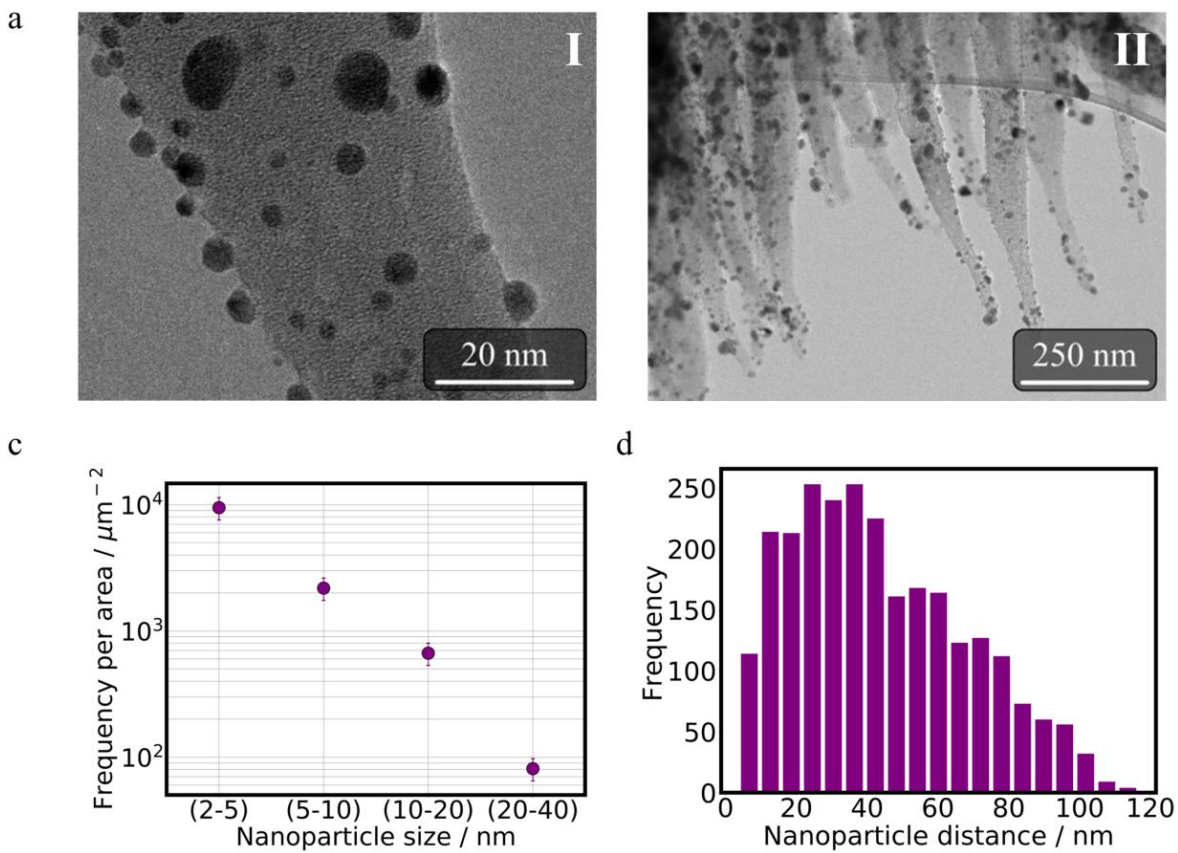

**Fig. 2: Particle size distribution of isolated Ag on silica hybrid nanowires fabricated with MS-ALD (20 ms TDMAS dose time, 6 s oxygen plasma, 225 cycles, substrate temperature: 120°C). (a) TEM images of Ag-silica**

**hybrid structures and (b) Size distribution of Ag nanoparticles per area. We determined the particle size distribution by counting classes of nanoparticles on defined nanowire surfaces for several TEM images using image processing. The number of particles was normalized to 1 µm² nanowire surface area. (c) Calculated nanoparticle distances between all particles from Fig. 2 (a-l).**

Fig. 2(a) shows transmission electron microscopy (TEM) images for Ag-silica hybrid structures at different magnifications (see supporting information Fig. S3-6 for additional TEM images). As we determined by selective area electron diffraction, the structures consist of silica nanowires with Ag nanoparticles (dark regions) randomly attached to the nanowires.[15] The nanoparticles are not entirely spherical in shape. In addition, some particles have merged to larger particles, leading to the distinct heterogeneity in the particle shapes. Fig. 2(b) shows the concentration of nanoparticles on the normalized surface as a function of the nanoparticle size. We could identify nanoparticles with diameters from 2 nm to 40 nm. However, the 3D structures mainly consist of small nanoparticles (2-10 nm): their concentration is 15 times larger than the concentration of larger particles (10-40 nm).

Individually, neither such small Ag particles (Fig. S8) nor Fresnel calculations (Fig. S9) on the initial material stack could adequately describe the high absorbance we could observe (see Fig. 1(d). Nevertheless, there have been several studies reporting on and modeling slight broadening effects on the scattering and the absorption bandwidth of structures comprised of small particles. For instance, Liu et al. observed a SPR peak shift of the extinction spectra from 425 nm to 470 nm for even larger Ag particles of 60 nm and 100 nm, respectively,[25] yet the extinction rate decreased markedly for wavelengths larger than 600 nm. Mertens et al. observed a redshift of the absorption caused by an anisotropy of the particle shape and an interparticle coupling for gap distances below 30 nm, so-called plasmon hybridization.[26] Lu et al. achieved an absorbance of

> 90 % in the range from 300-800 nm, depending on the surface filling factor with Ag particles deposited by magnetron sputtering.[27]

In our study, the Ag nanoparticles revealed anisotropic particle shapes (see Fig. 2(a)). Moreover, they were randomly distributed onto the silica nanowire surface and over the entire surface of the 3D structure. This led to observable gap distances which we also calculated using image processing.[28] The distances range from 1 nm up to 120 nm (see Fig. 2(c)). More details can be found in the SI.

Due to the statistical distribution of the used materials with irregular features of various sizes and geometries, the underlying optical effects caused by the resulting structures are manifold. In general, as described by Mie theory, the eigenmodes (dipolar, quadrupolar, etc.) of the plasmonic nanoparticles are expected to contribute to the overall optical signal.[24] Also, effects from morphological anisotropies are known, as well as wide-range frequency shifts along the visible spectrum arising from the electromagnetic coupling of near-touching particles. This was described by early theoretical studies utilizing discrete dipole-dipole approximations and later experimentally verified for Ag nanodisks in close proximity which exhibited strong coupling.[29,30] Additionally, a redshifting of the LSPR wavelength can also be observed, depending on the degree of embedding in the substrate[31] and the refractive index of the substrate.[32] Array-like and lattice-like superstructures also play a major role in the overall absorption and scattering properties. This is especially relevant in the case of very densely packed particles that exhibit short-range or long-range orders.[33–36]

In our study, a highly enhanced light extinction is caused by multiple scattering and absorption by the spatially distributed Ag nanoparticles in 3D structures with structure heights of more than 4 microns (see Fig. 3(a)). In addition, the strong coupling of plasmon modes between those particles that are close to each other (hybridization effect) also contributes markedly to the enhanced absorption over a large wavelength range. In other words, the increase in light extinction in broadband is caused by complex optical interactions with the spatially distributed Ag nanoparticles. Consequently, when we removed the Ag nanoparticles from the 3D hierarchical silica structures, the absorbance decreased dramatically (Fig. S2).

In order to examine the mentioned effects as a combined cause for the absorption spectra we observed, we simulated the optical response using an image-derived finite element method similar to the approach described by Trautmann et al.[37] A TEM image of an exemplary structure (see Fig. 3(b)) was used to remodel a domain of the fabricated structures for numerical calculations. We selected a representative, nanowire-like structure containing Ag nanoparticles of different sizes and shapes. We obtained the exact locations and sizes of the statistically distributed particles using image processing, determining the locations by calculating the object centroids. The actual particle size was then converted into a circle of equivalent area as shown in Fig. 3(c).

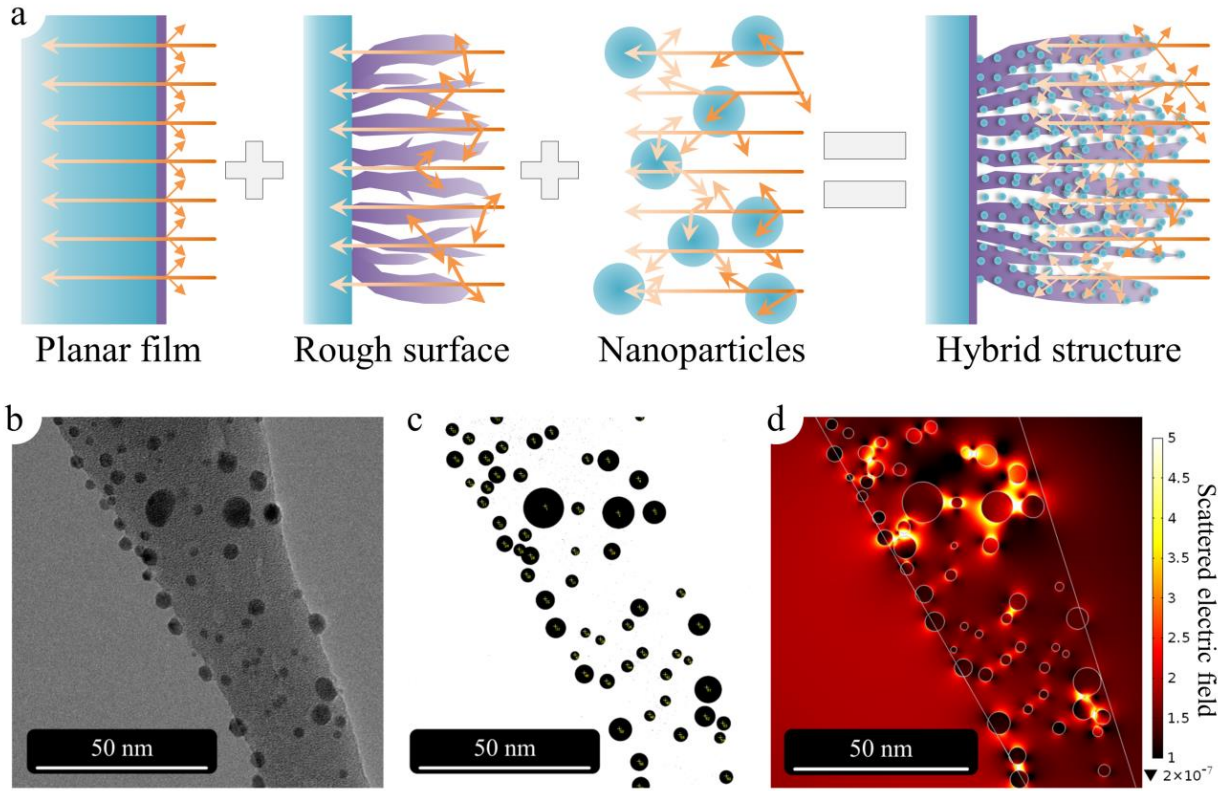

**Fig. 3:** (a) Scheme illustrating the concept of mixing different architectures and materials to generate high absorption structures. (b) TEM image illustrating the silica nanowire and Ag nanoparticles (darker regions), (c) binary image highlighting the position and the size of the Ag nanoparticles. (d) Corresponding scattered electric field calculated by finite element methods at an incident wavelength of 400 nm.

The scattered electric field at wavelength 400 nm as solved for the two-dimensional modeled domain by finite element methods is depicted in Fig. 3(d). The optical properties of the Ag particles have been determined using literature values.[38] We could observe both strong coupling and mode volume confinement between the individual particles as well as a diffuse scattering at various angles (Fig. S10). Note that the selected modeling approach is different from the mentioned hybridization effects as the particles in the simulation, in contrast to the morphologically anisotropic nanoparticles from the real 3D structure, were spherical in shape. However, the spectral contribution of the hybridized plasmon modes originating from anisotropic particles was modeled indirectly by the strong coupling of several near-touching individual spherical particles which exhibited an equivalent optical response.[39] This can be assumed as long as the preferential particle shape over the sample volume is spherical and no dominating morphological long-range order can be found. The spectra obtained by experiment and shown in Fig. 4(a) are expected to be a result of all statistically distributed Ag particle arrangements situated on the silica nanowires. We plotted two different samples with different average diameters and structure heights as shown in Table 1. As can be seen, with a higher number of cycles, structure heights increase; the average nanoparticle diameter, however, decreases.[15,19] Simultaneously, the high extinction range increases as the number of cycles increases, too. However, the amount of silver remained unchanged in both samples, as there is no further deposition of silver during the 3D evolution process. The observed differences in extinction are due to the differences in height and the diverse nanoparticle diameters. So, the amount of small nanoparticles has increased for structures fabricated at a higher number of cycles. In addition, the smaller nanoparticle diameters led to the absorption having a greater impact than the scattering had, just as predicted by Mie.[24]

| Number of cycles | Structure height | Nanowire diameter | Nanoparticle diameter |
|---:|---:|---:|---:|
| 135 | 7.2 µm ± 0.5 µm | 152.9 nm ± 29.2 nm | 18.9 nm ± 15.2 nm |
| 225 | 9.5 µm ± 1.1 µm | 168.0 nm ± 29.5 nm | 11.8 nm ± 7.8 nm |

**Table 1: Metric properties of the samples illustrated in Fig. 4(a) as a function of the number of MS-ALD cycles, TDMAS dose time: 20 ms, oxygen plasma dose time: 6 s, Ag film thickness: 300 nm.**

With the structural diversity of the composite dendrites and the given variety of orientations, we could markedly enhance the surface area with homogeneously covered Ag particle film on the macroscopic level. The incoming light interacts with different features of the structure, dependent on their size, and is scattered diffusively. We observed this light trapping mediated by surface plasmons over a broad wavelength range for these Ag-silica composite structures. The modeled domain for simulation consisted of 56 particles with an average size of $\mu_0 = 4$ nm as shown in Fig. 2(a-I). We evaluated the expected spectral position and bandwidth of the total sample by upscaling the modeled domain (a bunch of Ag nanoparticles on a nanowire with a certain diameter value as shown in Fig. 3(c)) at an incident angle $\vartheta = 70°$. The scaling coefficients $\mu_s/\mu_0$ given in Fig. 4(b) represent a distribution of particles sizes from the defined classes in Fig. 2(b). The broadness of the extinction band increases dramatically with both scaling coefficient and particle size. These results are calculated from one bunch of Ag nanoparticles (Fig. 3(c)) with different scaling coefficients. In the real structures, there are countless bunches of Ag nanoparticles of different particle sizes. Somehow, the excellent broadband absorption is resulted from the complex optical behaviors (multiple scattering and absorption) in the spatially distributed Ag nanoparticles. Compared to a flat compact Ag thin film, the 3D hierarchical structures have a much larger penetration depth of light,[40] leading to the complex optical behavior described above.

As can be seen in Fig. 4(b), the simulated spectra of a single nanowire without nanoparticles have rather low extinction values. In contrast to that, extinction for the nanowire coated with nanoparticles is very high. This highlights the crucial role of embedded nanoparticles as extinction centers; yet the extinction was still quite narrow-banded in comparison to the experimental results (Fig. 4(a)). Substantial forward-scattering was found for an MS-ALD nanowire in comparison to the same particle ensemble embedded in air ($n_m = 1.0$), which can be explained by the angular distribution of scattered power by these structures as shown in Fig. 4(d). The effective absorption of the incoming radiation is amplified by preferential scattering along the glass nanowires further into the absorbing material. Nevertheless, Fig. 4(d) also shows a rate of scattering in other directions for a MS-ALD nanowire that cannot be neglected. So, the extinction can be increased remarkably by adding a second MS-ALD nanowire in close vicinity to the first MS-ALD nanowire (Fig. 4(c)). The second nanowire provides more scattering centers due to an increased number of silver nanoparticles and leads to a more unidirectional extinction in comparison to the extinction of a single nanowire based on forward-scattering. Hence, the total optical far-field response of the broadband absorber was approximated from the near-field responses of individual features as present in the structure as follows: (1) by summation of the calculated extinction from the polydisperse particle distribution present in the Ag-silica composite structures (Fig. S11), (2) by the presence of multiple scattering processes emphasized by the high density of particles and material boundaries and (3) by preferential scattering into the structures synthesized with MS-ALD.

Despite the restriction to a two-dimensional domain, the calculated spectral bandwidths as obtained by this scaling method are in good agreement with the broadband absorption achieved in experiment. Thus, the optical signals of such samples with randomly and irregular oriented, yet homogeneously distributed nanostructures can easily be estimated using a representative, morphological domain of the whole structure.

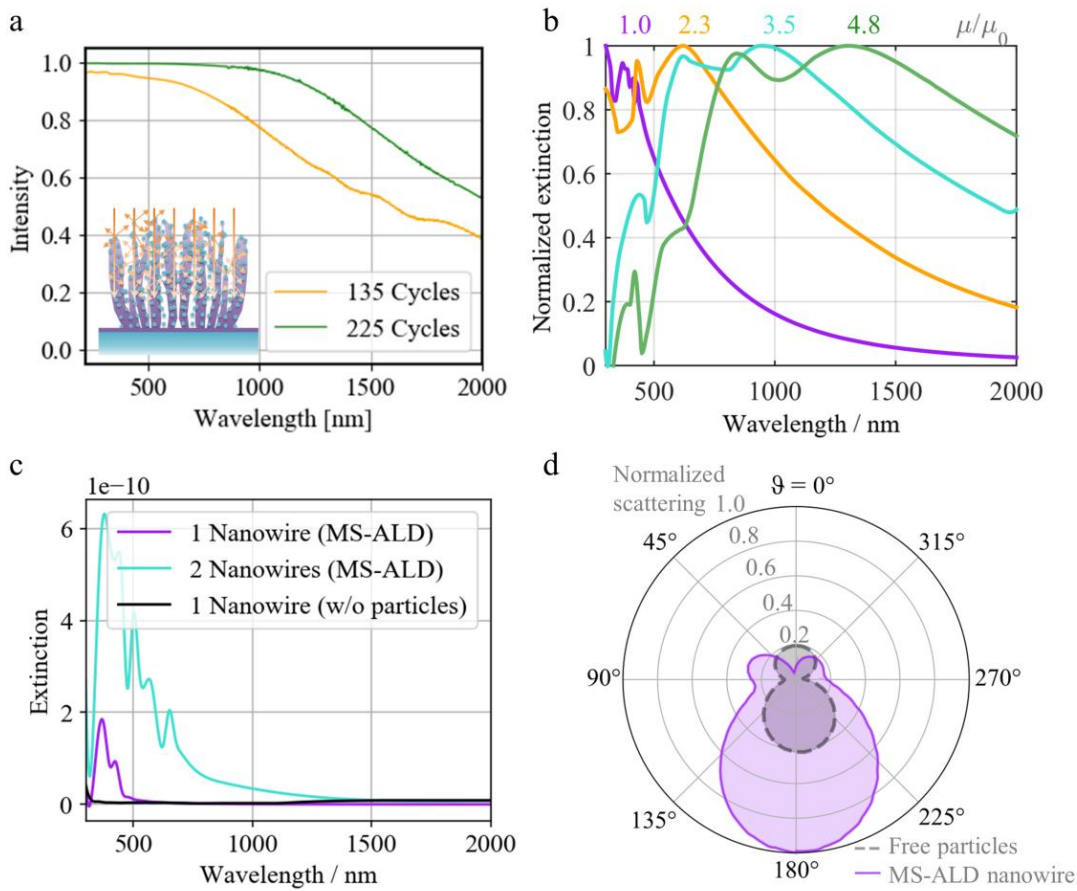

**Fig. 4: (a)** Experimentally obtained absorption spectra from two different samples fabricated with 135 and 225 MS-ALD cycles, respectively. Inset: illustration of multiple scattering processes at boundaries and nanoparticle positions **(b)** Calculated extinction (at $\vartheta = 70°$) for different coefficients $\mu_s/\mu_0$ describing the upscaling factor in respect to the initial simulation domain. The extinction spectrum of the initial domain (1.0) – purple, and other scaling coefficients (2.3; 3.5; 4.8) are shown to assess spectral position and bandwidth. **(c)** Calculated extinction spectra (at $\vartheta = 0°$) of a single MS-ALD nanowire, double MS-ALD nanowire and glass nanowire without Ag-particles as reference (see Fig. S 12 in SI). **(d)** Angular distribution of the scattered power for an individual MS-ALD nanowire and a particle ensemble in air (n = 1.0). Increased directionality causes preferential scattering into the absorbing material.

The agreement between simulations and experimental data indicates that the experimentally obtained spectra result from the randomly distributed Ag-particle arrangements located on the silica nanowires. Thus, we would be able to increase absorption in both bandwidth and intensity by increasing the amount of various particles of different shapes into the 3D hierarchical silica scaffold. In experiment, a second supercycle (Ag deposition, MS-ALD synthesis) deposited on top of the structure indeed caused an increase of the randomly distributed Ag-particle arrangements. This higher number of nanoparticles led, in turn, to an expansion of the absorption range from 200 nm up to 1400 nm. Going one step further, the deposition of a third supercycle led to a further increase of absorbance (> 98 %) up to a maximum of 2500 nm due to an increased amount of Ag nanoparticles embedded in the 3D structure. This agrees with theories of multi-scattering processes enabled by a larger penetration depth and preferential scattering into the absorbing material. Nevertheless, further studies have to be conducted to address the polarization dependency and angular properties of the spectral characteristics with incident light to benchmark the described structures under various conditions and to render them comparable with other structures.[14,41,42]

**Author contribution**

M.Z. and A.D. contributed equally to this work. The transmittance/reflectance measurements, design of experiments and the sample fabrication were conducted by M.Z. The simulations and design of experiments were conducted by A.D. The TEM measurements were conducted by K.P. and M.Z.; A.D. and D.W. have prepared the manuscript. All authors discussed the experimental as well as the theoretical results and contributed to the writing of the manuscript.

**Acknowledgement**

Partial funding of Deutsche Forschungsgemeinschaft (DFG, Grant SCHA 632/24) is gratefully acknowledged. F. Langenhorst thanks the Deutsche Forschungsgemeinschaft for funding the TEM via the Gottfried Wilhelm Leibniz programme (LA 830/14-1).